\documentclass[a4paper, 10pt]{article}

\usepackage{enumerate}
\usepackage{amsmath}
\usepackage{algorithm}
\usepackage{algorithmic}
\usepackage{graphics}
\usepackage{array}
\usepackage{url}
\usepackage{graphicx}

\newcolumntype{C}{ >{\centering\arraybackslash} m{8mm}}
\newcolumntype{D}{ >{\centering\arraybackslash} m{8mm}}
\newcolumntype{R}{ >{\centering\arraybackslash} m{6mm}}
\newcolumntype{Q}{ >{\centering\arraybackslash} m{5mm}}
\newcolumntype{E}{ >{\centering\arraybackslash} m{20mm}}

\usepackage[a4paper]{geometry}
\title{A Parallel Simulated Annealing Approach for the Mirrored Traveling Tournament Problem}
\date{}

\begin{document}

\maketitle

\begin{abstract}

The Traveling Tournament Problem (TTP) is a benchmark problem in sports scheduling and has been extensively studied in recent years. The Mirrored Traveling Tournament Problem (mTTP) is variation of the TTP that represents certain types of sports scheduling problems where the main objective is to minimize the total distance traveled by all the participating teams. In this paper we test a parallel simulated annealing approach for solving the mTTP using OpenMP on shared memory systems and we found that this approach is superior especially with respect to the number of solution instances that are probed per second. We also see that there is significant speed up of 1.5x - 2.2x in terms of number of solutions explored per unit time. 

\end{abstract}

\section{Introduction}
\label{intro}
Mathematically optimized schedules have huge practical roles as they often have a large impact both economically and environmentally and one area where they provide significant results is in sports league scheduling. Professional sports leagues exists as big businesses all over the world and many of these popular leagues are of huge economic importance due to the vast amounts of revenue they generate. While economic importance is one of the reasons the Traveling Tournament Problem (TTP) has received much attention in recent years, one other major reason is the extremely challenging scheduling problems they generate. In fact while the general complexity of TTP is still an open question, some instances of it have been proved to be NP-complete \cite{Westphal, Rishi}. \newline

The TTP was introduced by Easton et.al in 2001 \cite{Easton} and the problem, given the number of teams $n$ (even) and the pairwise distance between their home venues, is concerned with arriving at a schedule for a double round robin tournament that minimizes the sum of the distances traveled by all the participating teams. While arriving at an optimized schedule, $S$, for the double round robin tournament, TTP places two additional constraints on the schedule called the AtMost and the NonRepeat constraint. The AtMost constraint mandates that each team must play no more than $k$ ($k$ is usually taken as 3) consecutive matches at home or away and the NonRepeat constraint states that two teams should not play each other in consecutive rounds. \newline

In this paper we consider an important variant of the TTP called the Mirrored Traveling Tournament Problem (mTTP). mTTP was introduced by Ribeiro and Urrutia in \cite{mTTP} and here in place of the NonRepeat constraint we have a Mirror constraint. The Mirror constraint requires that the games played in round $r$ are exactly the same as those played in round $r + (n - 1)$, for $r = 1, 2, \cdots, n - 1$, with reversed venues. While there have been many attempts at arriving at optimized schedules for both the TTP and mTTP \cite{Anag, mTTP, Lim, Easton}, here we suggest a parallel simulated annealing approach for solving the mTTP and we show that this approach is superior especially with respect to the number of solution instances it can probe per unit time. Additionally, based on an implementation on OpenMP, we also show that there is significant speed up of 1.5x - 2.2x in terms of number of solutions it can explore per unit time. 

\section{Methodology}

Simulated Annealing is a local search meta-heuristic used to address global
optimization problems, especially when the search space is discrete. The name
comes from the process of annealing in metallurgy which involves the heating and
controlled cooling of a metal to increase the size of its crystals and to reduce
their defects. If the cooling schedule is sufficiently slow, the final
configuration results in a solid with superior structural integrity which in
turn represents a state with minimum energy.Simulated annealing emulates the physical process described above and in this
method, each point $s$ of the search space is analogous to a state of some
physical system, and the function $E(s)$ that is to be minimized is analogous to
the internal energy of the system in that state.

In the following subsections, we explain the serial version of mTTP and then we discuss the parallelization of this algorithm.

\subsection{The SA algorithm for mTTP} \label{sa}

The simulated annealing algorithm starts with an initial random schedule, $S$ and at each basic step it probabilistically decides between making a transition to a schedule $S'$ in its neighborhood, or staying at $S$. The neighborhood of a schedule $S$ is defined as the set of all schedules that can be generated by applying any one of the $5$ five moves : swap-teams, column-swap, row-swap, swap-rounds, interchange-home-away. These $5$ moves are the same as those suggested in \cite{Anag}. 

%\begin{enumerate}
%\item Swap-Teams()
%\item Col-Swap()
%\item row-swap()
%\item Swap-rounds()
%\item Interchange-Home-Away()
%\end{enumerate}  

Once the neighbouring schedule $S'$ is determined, the probability of making the transition to the new configuration $S'$ is dependent on the on the variation, $\Delta$, in the objective function produced by
the move. The system moves to $S'$ with a probability 1 if $\Delta < 0$. If
$\Delta > 0$, then the transition to the new state $S'$ happens with a
probability $\exp({-\Delta}/{T})$. The rationale behind this is that, here as the temperature decreases over time
the probability, $\exp({-\Delta}/{T})$, of accepting non-improving solutions 
decreases.

\subsection{The Parallel SA algorithm for mTTP} \label{psa}

In order to overcome the restrictive nature of the serial SA algorithm presented in \ref{sa} in terms of the number of solutions being explored, in this paper we explore the possibility of parallelism in the SA algorithm. Since the nature of the SA algorithm allows only for work level parallelism, we the exploit work level parallelism offered by shared memory multi core CPU's using openmp (omp) threads and we present the parallel simulated annealing algorithm ( PSA(T) ) below, where T is the number of threads used. The main rationale behind choosing this model comes from the intuition that as the number of threads increases the solutions explored by them, collectively, will be significantly larger and hence would help us in obtaining the optimal solutions faster.
\newpage
\begin{algorithm}[htbp]
\label{alg2}
\caption{: PSA(T)}
\begin{algorithmic}[1]
\STATE \textbf{do in parallel} for each thread $1, 2 \cdots T$
\STATE start with a random schedule $S$ 
\STATE curr$\_$dist $=$ best$\_$dist $=$ distance($S$)
\STATE curr$\_$schedule $=$ best$\_$schedule $= S$
			
\STATE initialize n$\_$iterations, $T_{\text{initial}}$, $T_{\text{final}}$ and
$\alpha$	
\STATE set count$\_$itr $= 0$
	
\WHILE{(count$\_$itr $<$ n$\_$iterations)}	
	\STATE temp$\_$curr $=$ $T_{\text{initial}}$ 	
	\STATE temp$\_$end =  $T_{\text{final}}$
		
	\STATE curr$\_$dist = best$\_$dist;
	\STATE curr$\_$schedule = best$\_$schedule
		
	\WHILE{(temp$\_$curr $>$ temp$\_$end)}
			
	\STATE S' = select$\_$random$\_$schedule()
	\STATE total$\_$dist $=$ distance(S')
			
	\STATE $\Delta =$ total$\_$dist $-$ curr$\_$dist
			
	\IF{($\Delta < 0$ \OR $\exp(-\Delta /$ temp$\_$curr) $>$ random()) }  
				
		\STATE curr$\_$dist $=$ total$\_$dist
		\STATE curr$\_$schedule $=$ S' 
			
		\IF{(total$\_$dist $<$ best$\_$dist)} 
			\STATE acc $=$ check$\_$schedule()
					
				\IF{(acc is \TRUE)}
					
					\STATE best$\_$dist $=$ curr$\_$dist
					\STATE best$\_$schedule $=$
curr$\_$schedule
					
				\ENDIF
					
		\ENDIF
				
	\ENDIF
			
	\STATE temp$\_$curr $=$ temp$\_$curr $ * \alpha$;
			
\ENDWHILE
		
\STATE count$\_$itr++
		
\ENDWHILE
\STATE \textbf{end do in parallel}
\STATE synchronizeThreads()
\STATE Pick least distance schedule from all the threads

\end{algorithmic}
\end{algorithm}

\section{Computational Experiments and Results} \label{results}
The proposed parallel simulated annealing algorithm was tested on a number of mTTP instances given in \cite{Web} and it was seen that this algorithm, in addition to finding optimized solutions for these instances (all of which were within 10\% of the known lower bounds), was superior especially in terms of the number of solutions that could be explored in a second. This is particularly significant since one of the main objectives of a simulated annealing approach is to explore as much of the solution space as possible. Figure 1 demonstrates the variation in the number of solutions explored using the serial SA, PSA(2) and PSA(4) for instances NL06, NL08, CIRC08, NL10 and CIRC10. Figure 2 provides the corresponding speed up graph for these instances. It is evident from the figure that a significant speedup of upto 2.2X was achieved. 

\begin{figure}[h]
\begin{minipage}{16pc}
\includegraphics[scale=0.35]{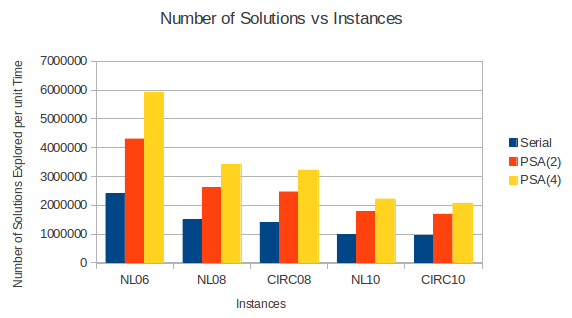} 
\caption{\label{fig4} Variations in the number of solutions explored.}
\end{minipage}\hspace{2pc}%
\begin{minipage}{16pc}
\includegraphics[scale=0.35]{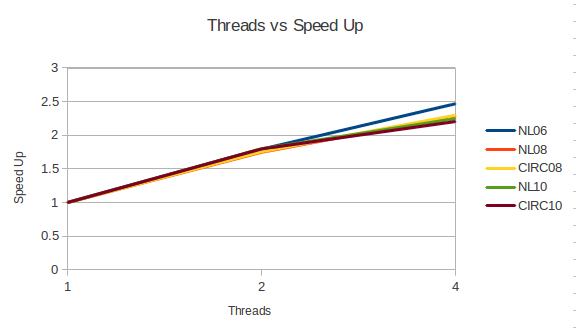}
\caption{\label{fig4} Threads Versus Speed Up of Annealing on mTTP.}
\end{minipage}
\end{figure}

\section{Conclusion}
Annealing belongs to class of sub optimal algorithms which depends heavily on randomization. In order to improve the solution, we need to explore more number of solutions at each basic step. The proposed parallel SA achieves this objective by utilizing multi core omp threads. Parallel SA will thus help in converging faster towards the optimal solution.

As for the future work, we plan to extend the proposed parallel version to incorporate synchronization and communication points between the threads for faster convergence towards the global optimum. We also plan to port the parallel SA to GPGPU's to achieve better performance using streaming multicore processors of NVIDIA's Compute Unified Device Architecture (CUDA) technology.

\end{document}